\documentclass[9pt,conference]{IEEEtran}
\IEEEoverridecommandlockouts

\usepackage{amsmath,amssymb,amsfonts}
\usepackage{algorithmic}
\usepackage{graphicx}
\usepackage{textcomp}
\usepackage{stfloats}
\usepackage{url}
\usepackage{verbatim}
\usepackage{graphicx}
\usepackage{cite}
\usepackage{multirow}
\usepackage{lscape,array}
\usepackage{booktabs}
\usepackage{pifont}
\usepackage{ragged2e}
\usepackage{enumitem}
\usepackage{xcolor}
\usepackage{hyperref}
\usepackage{balance}

\def\BibTeX{{\rm B\kern-.05em{\sc i\kern-.025em b}\kern-.08em
    T\kern-.1667em\lower.7ex\hbox{E}\kern-.125emX}}
\begin{document}

\title{
CA-MHFA: A Context-Aware Multi-Head Factorized Attentive Pooling for SSL-Based Speaker Verification
}

\author{
\IEEEauthorblockN{Junyi Peng}
\IEEEauthorblockA{\textit{Speech@FIT} \\
\textit{Brno University of Technology}\\
Czechia}
\and
\IEEEauthorblockN{Ladislav Mošner}
\IEEEauthorblockA{\textit{Speech@FIT} \\
\textit{Brno University of Technology}\\
Czechia}
\and
\IEEEauthorblockN{Lin Zhang}
\IEEEauthorblockA{\textit{Speech@FIT} \\
\textit{Brno University of Technology}\\
Czechia}
\and
\IEEEauthorblockN{Oldřich Plchot}
\IEEEauthorblockA{\textit{Speech@FIT} \\
\textit{Brno University of Technology}\\
Czechia}
\and
\IEEEauthorblockN{Themos Stafylakis}
\IEEEauthorblockA{\textit{Athens University of Economics and Business}\\ \textit{Omilia, Archimedes/Athena R.C.}\\
Greece}
\and
\IEEEauthorblockN{Lukáš Burget}
\IEEEauthorblockA{\textit{Speech@FIT} \\
\textit{Brno University of Technology}\\
Czechia}
\and
\IEEEauthorblockN{Jan Černocký}
\IEEEauthorblockA{\textit{Speech@FIT} \\
\textit{Brno University of Technology}\\
Czechia}
}

\maketitle
\begin{abstract}
Self-supervised learning (SSL) models for speaker verification (SV) have gained significant attention in recent years. However, existing SSL-based SV systems often struggle to capture local temporal dependencies and generalize across different tasks. In this paper, we propose context-aware multi-head factorized attentive pooling (CA-MHFA), a lightweight framework that incorporates contextual information from surrounding frames. CA-MHFA leverages grouped, learnable queries to effectively model contextual dependencies while maintaining efficiency by sharing keys and values across groups. Experimental results on the VoxCeleb dataset show that CA-MHFA achieves EERs of 0.42\%, 0.48\%, and 0.96\% on Vox1-O, Vox1-E, and Vox1-H, respectively, outperforming complex models like WavLM-TDNN with fewer parameters and faster convergence. Additionally, CA-MHFA demonstrates strong generalization across multiple SSL models and tasks, including emotion recognition and anti-spoofing, highlighting its robustness and versatility. \footnote{Code is available at \url{https://github.com/BUTSpeechFIT/wespeaker_ssl_public}}
\end{abstract}
\section{Introduction}
\label{sec:intro}

Large-scale self-supervised learning (SSL) speech models, such as Wav2vec \cite{baevski2020wav2vec}, HuBERT \cite{hsu2021hubert}, WavLM \cite{chen2022wavlm}, and their variants \cite{baevski2022data2vec}, have significantly advanced various speech-related tasks, including speech recognition (ASR) \cite{li2023parameter}, emotion recognition (ER) \cite{9747870}, deepfake detection \cite{wang22_odyssey}, and speaker verification (SV) \cite{chen2022large, peng2022parameter}. These SSL models are pre-trained on extensive speech datasets, which allow them to learn universal representations that are more robust to channel mismatch and noisy conditions compared to traditional acoustic features like Mel-frequency cepstral coefficients (MFCCs) and FBank, especially under low-resource scenarios \cite{peng23_interspeech, hsu2023low}.

The most common approach to adapting these general-purpose models to specific downstream tasks is by fine-tuning the entire pre-trained SSL model with a task-oriented back-end module using labeled downstream data. In the field of SV, since the outputs of SSL models are layer-wise frame-by-frame representations, the task-oriented module (known as the speaker extractor back-end) processes these features to generate utterance-level speaker embeddings, which are then used for speaker identity prediction. It typically includes a frame-level feature extractor, a pooling mechanism, and an utterance-level feature extractor. 
For example, in \cite{yang2021superb, chen2022large}, a weighted sum of layer-wise SSL outputs replaces traditional acoustic features as input to speaker extractors, such as the x-vector~\cite{snyder2018x} and the ECAPA-TDNN models~\cite{desplanques20_interspeech}. These SSL-based SV systems significantly accelerate convergence and boost performance compared to those based on MFCC features. 
To further explore the potential of SSL models, a lighter back-end, called multi-head factorized attentive pooling (MHFA) \cite{peng2022attention}, was proposed. MHFA employs two sets of weights to model different aspects of input features. This approach has been shown to outperform systems that integrate ECAPA-TDNN on the VoxCeleb dataset.

Although MHFA has achieved promising results on the SV task, it typically operates at the utterance level. By processing all frames simultaneously, it ignores the detailed relationships between surrounding frames. This lack of local context limits its ability to capture dynamic and fine-grained temporal dependencies across frames. Additionally, since pre-trained SSL models already possess strong frame-level modeling capabilities by leveraging self-attention mechanisms, incorporating complex and randomly initialized frame-level modules in the back-end can mislead the optimization process, leading to suboptimal performance \cite{aldeneh2024can}. Furthermore, studies on the generalizability of these lightweight back-end modules to other speech classification tasks, such as emotion recognition (ER) and spoofing detection, remain limited. This raises the concern that these modules might be over-optimized for a single task, such as SV \cite{zaiem23b_interspeech}.

To address the aforementioned challenges, in this paper, we aim to develop a lightweight framework considering context information for SV and can be applied to broader speech classification tasks and various SSL models.
The proposed framework, named context-aware multi-head factorized attentive pooling (CA-MHFA), utilizes both past and future speech frames as contextual information to extract utterance-level representations from SSL layer-wise features.
Specifically, similar to the self-attention mechanism used in MHFA~\cite{peng2022attention}, we employ two sets of learnable weights along with a linear layer to generate \emph{keys} and \emph{values}. 
Unlike MHFA, we introduce a set of learnable \emph{queries} that are grouped, allowing the model to focus on the surrounding frames and better capture contextual dependencies.
The attention weights are computed using convolution between the grouped \emph{queries} and \emph{keys}.
Finally, an attentive pooling layer aggregates all frames from each group, followed by a concatenation and linear layer to compute the speaker embedding. 
Moreover, we share the single key and value with the grouped queries to simplify the entire pipeline. 

The contributions of our work are as follows:

\begin{itemize}
    \item \textbf{A lightweight back-end module:} We propose a lightweight back-end, CA-MHFA, to extract representations that consider contextual information from nearby frames when calculating attention weights.
    \item \textbf{Compatible architecture:} Besides MHFA, our context-aware extension is compatible with other pooling methods, like mean pooling and attentive pooling~\cite{zhu18_interspeech}, making it extensible for future integration and improvements to existing ones.
    
    \item \textbf{State-of-the-art SV performance:} We achieve SOTA results on the VoxCeleb dataset using fewer model parameters. With the same pre-trained SSL model, the proposed system outperforms the WavLM-TDNN \cite{chen2022wavlm} and yields 0.42\%, 0.48\% and 0.96\% EER on Vox1-O, Vox1-E and Vox1-H, respectively.  
    \item \textbf{Strong generalization:} We demonstrate the effectiveness of the proposed lightweight back-end module across various speech classification tasks and SSL models, including SV, ER, and deepfake detection, following the SUPERB principle. This is evaluated by integrating with nine popular pre-trained SSL models, including Wav2Vec 2.0, HuBERT, Data2vec, and WavLM. 

\end{itemize}

\begin{figure*}
    \centering
    \includegraphics[width=0.99\linewidth]{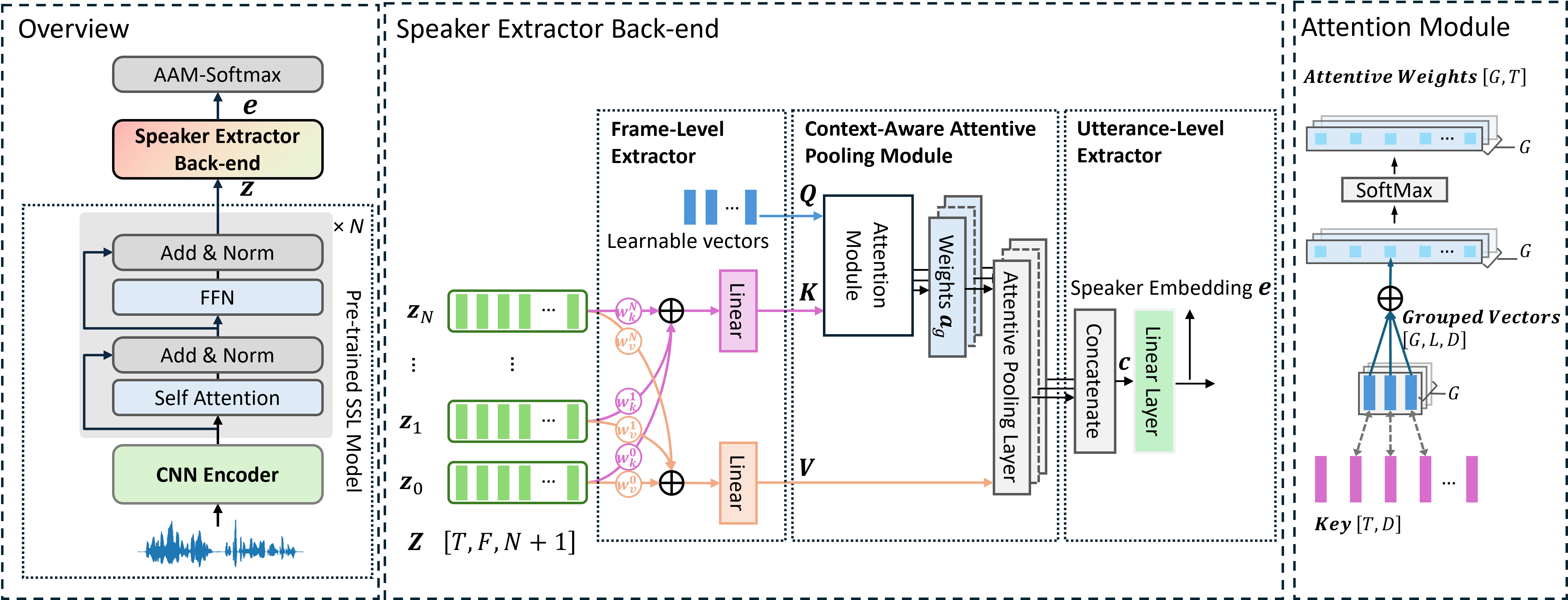}
    \caption{The architecture of the pre-trained model and attached lightweight speaker extractor back-end (context-aware multi-head factorized attentive pooling, CA-MHFA). During the fine-tuning, the SSL model and cascaded speaker extractor are jointly optimized.}
    \label{fig:enter-label}
\end{figure*}

\section{Context-aware Multi-head Factorized Attentive Pooling}
\label{sec:format}
In this section, we describe the proposed CA-MHFA in detail. As shown in Fig. \ref{fig:enter-label}, it consists of three main components: a frame-level extractor, a context-aware attentive pooling module, and an utterance-level extractor. 

\subsection{Frame-Level Extractor with Compression}

The goal of text-independent speaker embedding extractors is to capture speaker-related characteristics independent of content. 
However, SSL models are designed as general-purpose models, not specifically for speaker embedding extraction. As a result, the representations they extract often contain a mix of both speaker-related and content-related information \cite{yang2021superb}. Although the weighted sum method \cite{yang2021superb}, which uses learnable weights for each layer, can help mitigate this issue, it treats content and speaker information within the same layer equally. Consequently, when the method reduces the weights for layers rich in content-related information, it also unintentionally reduces the speaker-related information from those layers.
To effectively utilize and refine speaker-related information from all layers, the proposed CA-MHFA is designed to factorize SSL representations into multiple subsets, focusing on extracting speaker-specific features while minimizing interference from content.


To describe the proposed CA-MHFA, we use the \emph{query/key/value} abstraction for the attention mechanism~\cite{vaswani2017attention}. For a SSL model, we define outputs from its different layers as $\mathbf{Z}=\{\mathbf{z}_0,...,\mathbf{z}_N\}$, $\mathbf{z}_{n}\in\mathbb{R}^{T\times F}$, where $T$ is the dimention of the frame level, $F$ is the feature dimension of outputs from Transformer blocks, and $N$ denotes the number of Transformer blocks inside the SSL model. To construct multiple subsets, we employ two sets of normalized weights (factors) $\omega^k_{n}$ and $\omega^v_{n}$ along with a linear layer to separately aggregate and compress layer-wise outputs to produce the matrices of \emph{keys} $\mathbf{K}$ and \emph{values} $\mathbf{V}$, respectively:
\begin{equation}
\label{eq1}
\begin{split}
\mathbf{K} &= \left(\sum_{n=0}^{N} \omega_{n}^k\mathbf{z}_{n}\right)\mathbf{S}^k \\
\mathbf{V} &= \left(\sum_{n=0}^{N} \omega_{n}^v\mathbf{z}_{n}\right)\mathbf{S}^v,\\
\end{split}
\end{equation}
where $\mathbf{S}^k, \mathbf{S}^v \in \mathbb{R}^{F \times D}$ denote transformation matrices responsible for compressing the weighted frame-level features into keys and values, respectively, and $D$ represents the feature dimension after compression.

\subsection{Context-Aware Attentive Pooling}

To effectively model the information within the key flow, we introduce a set of global, input-agnostic, learnable vectors, denoted as \emph{queries} $\mathbf{Q}\in\mathbb{R}^{LG\times D}$. For contextual modeling, queries $\mathbf{Q}$ are divided into $G$ groups, with each group's vectors denoted as $\mathbf{q}^{g}=[\mathbf{q}_1^{g},..,\mathbf{q}_{L}^{g}]$, $\mathbf{q}^{g}\in\mathbb{R}^{L\times D}$, where $g\in[1, G]$ indexes the groups (heads) and each group $\mathbf{q}^{g}$ comprises $L$ trainable vectors. For computational efficiency, each of the groups $\mathbf{q}^g$ shares the same key and value. 

To model contextual information, we consider $R$ frames on either side of the current frame, resulting in a total of $L$ neighboring frames used to compute attention weights for the current frame. The current frame refers to the frame in focus during attention calculation, with neighboring frames providing additional context information. Here, $R = \text{Floor}((L-1)/2)$. The attention weights of each group $\mathbf{a}^g\in\mathbb{R}^{T\times 1}$ are further computed by $\mathbf{q}^g$ and $\mathbf{K}$ as follows:
\begin{equation}
\label{mhfa}
\begin{aligned}
a_{t}^{g} &= \frac{\exp\left( \frac{1}{L}\sum_{j=-R}^{R}\mathbf{q}_{j}^{g} \mathbf{k}_{t+j}^\top \right)}{\sum_{i=1}^{T} \exp\left( \frac{1}{L}\sum_{m=-R}^{R} \mathbf{q}_{m}^{g} \mathbf{k}_{i+m}^\top \right)}, \\
\end{aligned}
\end{equation}
where $a_{t}^{g}$ represents the attentive weight given to the current $t$-th frame after considering the surrounding frames $L$, and $\top$ denotes transpose. From an implementation point of view, this process can be viewed as a 2D convolution operation. Specifically, the input $\mathbf{K}$ consists of $D$ channels, and the output channels correspond to the number of groups $G$. The kernel size is $(L, D)$. In this way, $\mathbf{Q}$ is reshaped into grouped vectors and serves further as the $G$ convolutional kernels, each with the shape $(L, D)$.

\subsection{Utterance-Level Extractor}
Next, each group's speaker representations $\mathbf{c}^{g}\in\mathbb{R}^{1 \times D}$ are aggregated via attention weights $\mathbf{a}^{g}$ and then concatenated as:
\begin{equation}
\label{mhfa}
\begin{aligned}
\mathbf{c}^g &= \sum\nolimits_{t=1}^{T} \mathbf{a}_{t}^{g} \mathbf{v}_t, \\
\mathbf{c} &= \mbox{concat} \left(\mathbf{c}^1,...,\mathbf{c}^G\right),
\end{aligned}
\end{equation}
where $\mathbf{c}\in\mathbb{R}^{1 \times GD}$ aggregates the speaker representation at the utterance level from all subspaces (heads). Finally, the prediction of the speaker label is performed by passing the representation through a subsequent linear layer, $L_2$ normalization, and a classification layer, which is needed only during training. During testing, the output of the $L_2$ normalized linear layer is considered as the \emph{speaker embedding}~
$\mathbf{e}$.

\subsection{Generalization of Other Works}
The CA-MHFA extends the MHFA framework by introducing flexibility in the attention mechanism. When we set $L=1$, CA-MHFA simplifies to the original MHFA as presented in \cite{peng2022attention}. 

In scenarios utilizing average pooling, setting $\mathbf{q}_l^g$ to a constant non-trainable zero vector implies that the similarity of each frame to every group query equals zero. After softmax normalization, this results in equal weights of $1/T$ for each frame feature, in this case, CA-MHFA degenerates to average mean pooling.

Compared to self-attentive pooling \cite{zhu18_interspeech}, setting $G=1$ implies that $\mathbf{Q}$ comprises a single group that contains only one D-dimensional trainable query. In this configuration, only a single set of weights is computed, making CA-MHFA equivalent to self-attentive pooling.

\begin{table}[t]
\centering
\caption{Hyper-parameter analysis of CA-MHFA with different context lengths $L$ and numbers of heads $G$ in terms of EER (\%) on VoxCeleb dataset. $\dag$ denotes the system applying convolution in both \emph{keys} and \emph{values}.}
\label{tab:my-table}
\scalebox{0.9}{
\begin{tabular}{lcccccc}
\toprule
Methods & \#Param & \#Context & \#Head & Vox1-O & Vox1-E & Vox1-H \\
\midrule
MHFA    &0.72M         & 1              & 16     &0.79        &0.85        &1.71        \\
MHFA    &1.25M         & 1              & 32     &0.79        &0.81        &1.61        \\
MHFA    &2.30M         & 1              & 64     &0.76        &0.79        &1.58        \\
\midrule
CA-MHFA &         1.25M & 3              & 32     & 0.76        & 0.76       &  1.54      \\
CA-MHFA &         1.26M & 5              & 32     & 0.73       &  0.76      &   1.53     \\
CA-MHFA &         2.30M & 3              & 64     & 0.73       &  0.74      &   1.49     \\
CA-MHFA &         2.31M & 5              & 64     &  \textbf{0.69}      &  0.74      &  1.47     \\
CA-MHFA &         2.36M & 9              & 64     &  0.70      &  \textbf{0.72}      &  \textbf{1.45}     \\
CA-MHFA &         2.41M & 17              & 64     & 0.69      &  0.74      &  1.47     \\
\midrule
CA-MHFA$\dag$ & 2.32M & 5 & 64 & 0.72 & 0.76 & 1.52\\
\bottomrule
\end{tabular}
}
\end{table}

\begin{table}[t]
\centering
\caption{Performance comparison with the proposed CA-MHFA with other full fine-tuned SSL-based systems, as well as speaker extractors trained from scratch. \textbf{ET.} denotes the ECAPA-TDNN model. $\dag$ denotes wespeaker's implementation. $\ddag$~represents our implementation.}
\label{tab:exp3-sota}
\scalebox{0.77}{
\begin{tabular}{lcccccc}
\toprule
\multirow{2}{*}{Methods} & \multirow{2}{*}{FLOPs} &\multirow{2}{*}{\#Param}& \multicolumn{1}{c}{Vox1-O} & \multicolumn{1}{c}{Vox1-E} & \multicolumn{1}{c}{Vox1-H} \\
\cmidrule(lr){4-4} \cmidrule(lr){5-5} \cmidrule(lr){6-6}
& & &EER & EER & EER \\
\midrule
ECAPA-TDNN (ET.)~\cite{desplanques2020ecapa} & 1.04G & 6.19M & 0.90  & 1.11  & 2.32  \\
ResNet221~\cite{wang2023wespeaker} & 21.29G & 23.79M	& 0.50  & 0.67  & 1.21  \\
ResNet293~\cite{wang2023wespeaker} & 28.10G &28.62M	& 0.44  & 0.65  & 1.18  \\
\midrule

WavLM\_BASE\_Plus + ET.~\cite{chen2022wavlm} & - & 94M + 6M & 0.84  & 0.92  & 1.75  \\
NEMO Large + MFA~\cite{cai2023leveraging} & - & 130M & 0.48  & 0.71  & 1.54  \\
Conformer + MHFA~\cite{peng23_interspeech} & - & 172M + 2.3 M & 0.65  & 0.93  & 1.86  \\
UniSpeech-SAT\_Large + ET.~\cite{chen2022large} & - & 316M + 6M & 0.53  & 0.56  & 1.18  \\
WavLM\_Large + ET. ~\cite{chen2022wavlm}  & - & 316M + 6M & 0.38  & 0.48  & 0.98  \\
WavLM\_Large + ET. ~\cite{chen2022wavlm}$\dag$  & - & 316M + 6M & 0.41  & 0.55  & 1.11  \\
WavLM\_Large + MHFA ~\cite{peng2022attention}$\ddag$ & 25.79G & 316M + 2.3M & 0.55  & 0.59  & 1.24   \\
\midrule
\midrule
WavLM\_Base\_Plus + CA-MHFA   & 11.05G & 94M + 2.3M & 0.70  & 0.72  & 1.45     \\
\quad + LMF/QMF   & 11.05G & 94M + 2.3M & 0.59 & 0.65  & 1.30    \\
WavLM\_Large + CA-MHFA  & 25.79G & 316M + 2.3M & 0.55  & 0.62  & 1.18   \\
\quad + LMF/QMF  & 25.79G & 316M + 2.3M & 0.42 & 0.48 & 0.96  \\
\bottomrule
\end{tabular}
}
\end{table}

\section{Experiments}
\subsection{Setup}\label{sec:exp:setup}
\subsubsection{Two Types of Evaluation} We explored two types of evaluation: 1) Full fine-tuning condition: Both the SSL and back-end modules are trainable during fine-tuning. 2) SUPERB-style: We follow the SUPERB principle \cite{yang2021superb} and only update the back-end module during the fine-tuning stage, while keeping the SSL model frozen.

\subsubsection{Datasets} We first focus on exploring the efficiency of the lightweight CA-MHFA for the SV task, involving two of the aforementioned evaluation methods: 1) For full fine-tuning conditions, to ensure we have enough data for finetuning SSL model, we use the development set of VoxCeleb2 for training, then use \textit{Vox1-O}, \textit{Vox1-E}, and \textit{Vox1-H} trials for evaluating. 2) For SUPERB-style, the training set is VoxCeleb1 and the evaluation set is \textit{Vox1-O} following \cite{yang2021superb}. We also explored the generalizability of CA-MHFA to two other speech classification tasks based on the SUPERB-style evaluation: 
For R, the models are trained and evaluated on the IEMOCAP corpus~\cite{busso2008iemocap}, which consists of approximately 12 hours of recordings in five sessions. Regarding deepfake detection, 
we use the ASVspoof 2019 LA~\cite{wang2020asvspoof} dataset.

\subsubsection{Implementation details} For all three aforementioned tasks, we utilize two scales of pre-trained SSL models listed in Table \ref{tab:exp1_SUPERB}: 1) The \emph{BASE} models, consisting of a CNN encoder and 12 layers of Transformers with approximately 94M parameters. The dimension of the Transformer output $F$ is 768. 2) The \emph{Large} models, which include 24 layers of Transformers with approximately 316M parameters. In the full fine-tuning configurations, the extracted speaker embedding dimension is 256. For the SV task under full fine-tune condition, we employ the AAM-softmax \cite{deng2019arcface} loss function with a margin of 0.2 and a scaling factor of 32. During gradient updates, the gradients of the SSL model are scaled by 0.1. To further boost performance, we adopt large margin tuning \cite{thienpondt2021idlab} with longer (5-second) segments and a margin of 0.5 for an additional 3 training epochs. The initial learning rate is set to 1e-4 and decreases to 1e-6 by the 10th epoch using the AdamW optimizer. All fine-tuning datasets are augmented with  with MUSAN \cite{snyder2015musan} and room impulse responses.  In addition, we used speaker-wise adaptive score normalization and
Quality Measure Functions~\cite{thienpondt2021idlab} to calibrate the scores. 

When evaluating the generalizability of the model across different tasks and upstream models, we adopt the SUPERB-style evaluation: For SV, all systems use the AM-softmax loss function. To keep consistent with the x-vector implementation within the SUPERB, the speaker embedding dimensions of ECAPA-TDNN, MHFA, and CA-MHFA, are set to 512. The MHFA model uses~8 heads, while the CA-MHFA model uses 8 heads with a context length of 9. For ER, cross-entropy (CE) loss is employed for optimization, and mean pooling is performed via a weighted sum across layer-wise features, followed by mean pooling along the time dimension and a final linear layer. For deepfake detection, CE loss is used, with a learning rate set to 1e-4 using the Adam optimizer, following \cite{wang22_odyssey}. For the back-end module, we use the weighted sum with LLGF (LCNN-BLSTM) when MHFA is not utilized, as this structure has shown the best performance when the SSL model is frozen \cite{wang22_odyssey}. No augmentation methods are applied.

\subsubsection{Performance Metrics}
For SV and deepfake detection, equal error rate (EER) is employed to measure the performancesER, we use a leave-one-session-out 5-fold cross-validation to report averaged accuracy.

\subsection{Hyper-Parameter Analysis of CA-MHFA}
We present here a hyper-parameter analysis of CA-MHFA for the SV task under the full fine-tuning evaluation. We focus on analyzing different context lengths ($L$) and numbers of heads ($G$) in terms of EER (\%) on the VoxCeleb dataset. WavLM BASE Plus is used as the pre-trained SSL model~\cite{chen2022wavlm}.
The results are shown in Table \ref{tab:my-table}.


It is observed that increasing the context length and the number of heads improves the performance of CA-MHFA, especially on Vox1-E and Vox1-H. This may be due to the fact that while the SV task does not require extensive context, considering adjacent surrounding frames can be beneficial. 
Compared to MHFA, CA-MHFA consistently outperforms MHFA in most configurations. 

We also compared systems that applied convolution operations to both the \emph{keys} and \emph{values} branches, as shown in the last row of Table~\ref{tab:my-table}. Specifically, the \emph{values} branches used 1D convolution operations with a kernel size of 5, where the input and output channels were set to $D$. The results show that this configuration performs worse compared to systems that applied convolution only to the \emph{keys} branch.
This suggests that contextual information may be more beneficial for the content-related subspace than for the speaker-related subspace. One possible explanation is that context in the content subspace helps capture the speaker's speaking style, such as the structuring of words, while the speaker subspace focuses on features that are invariant to the speaker's identity and, thus, requires less contextual attention.

\begin{table*}[htbp]
\caption{Comparison of different frozen upstream models across three downstream tasks: SV, ER, and deepfake detection with varying back-end models following SUPERB principle.}
\begin{center}
\label{tab:exp1_SUPERB}
\scalebox{0.95}{
\begin{tabular}{l|cccc|ccc|ccc}
\toprule
\multirow{2}{*}{Upstream} & \multicolumn{4}{c|}{SV EER(\%)$\downarrow$}   & \multicolumn{3}{c|}{Emotion Recognition ACC(\%)$\uparrow$}  & \multicolumn{3}{c}{Deepfake Detection} EER(\%)$\downarrow$ \\
                          & x-vector & ECAPA-TDNN & MHFA & CA-MHFA & MeanPooling \cite{yang2021superb} & MHFA & CA-MHFA & LLGF & MHFA & CA-MHFA\\
\midrule
Wav2vec2.0 BASE         & 5.43    & 3.56   & 3.11    & 3.07   & 60.55  &62.90  & 63.13   & 2.03 & 0.73 & 0.52 \\
HuBERT BASE              & 5.65    & 3.29  & 2.78     & 2.82  & 60.73  &62.35 & 63.96   & 2.39 & 1.19 & 1.22 \\
Data2vec BASE            & 6.39    & 3.97  & 3.38     & 3.26  & 65.43  &65.93 & 65.52  & 3.83 & 1.53 & 1.37  \\
WavLM BASE               & 4.84    & 3.04  & 2.45    & 2.41  & 62.48  & 65.06 & 67.64  & 2.13 & 0.93  & 0.73  \\
WavLM BASE Plus           & 4.10    & 2.67  & 1.87     & 1.79  & 66.72  &66.58  & 66.45  & 3.64 & 0.20 & 0.30 \\
\midrule
Wav2vec2.0 Large          & 6.06    & 2.86    &2.72   & 2.64   & 62.76  & 63.59 & 64.42  & 0.83 & 0.68 & 2.39  \\
HuBERT Large              & 6.02    & 2.71  &2.43     & 2.34  & 62.48  & 63.72 & 64.97 & 2.30 & 0.75 & 0.67 \\
Data2vec Large            & 7.62    & 3.11    &2.59   & 2.62  & 64.97  & 64.88 & 64.79 & 4.26 & 1.41 & 1.32 \\
WavLM Large               & 4.87    & 2.17    &1.78   & 1.77   & 67.92 & 69.72  & 71.52  & 1.54 & 2.23  & 1.21  \\
\bottomrule
\end{tabular}}
\end{center}
\end{table*}

\subsection{Comparison with State-of-the-art SV Systems}
Under the optimal hyperparameters identified in the previous section ($G=64$, $L=9$), we compared the proposed CA-MHFA with other state-of-the-art SV systems, including both SSL-based models and speaker extractors trained from scratch, as shown in Table \ref{tab:exp3-sota}.

It is observed that WavLM\_Large combined with CA-MHFA outperforms the SOTA model trained from scratch like ECAPA-TDNN and deep ResNet models (e.g., ResNet221 and ResNet293) with fewer training steps (23 epochs vs. 150 epochs).
Additionally, despite the WavLM\_Large model having a much larger number of parameters than ResNet293, the FLOPs are comparable when using 2 seconds of speech as input. This indicates that after convergence, the inference speed of WavLM\_Large + CA-MHFA is comparable to that of ResNet293. Furthermore, when comparing SSL models of similar scales, WavLM + CA-MHFA consistently outperforms other SSL-based SV systems with a lighter speaker extractor back-end. For example, under the same WeSpeaker framework, the proposed CA-MHFA outperforms ECAPA-TDNN systems when using both WavLM\_Large and WavLM\_Base\_Plus.

\subsection{ CA-MHFA on Other Tasks}
In this section, we explored the generalizability of the proposed CA-MHFA across three downstream tasks: SV, ER, and deepfake detection. The evaluation follows the SUPERB-style as we introduced in section \ref{sec:exp:setup}.
The results are summarized in Table~\ref{tab:exp1_SUPERB}.

In general, the proposed CA-MHFA consistently improves performance in multiple tasks and upstream models over baseline systems and MHFA. Specifically, for SV, when using the same SSL features as input, CA-MHFA consistently outperforms the x-vector and ECAPA-TDNN back-ends with fewer parameters (2.3M vs. 9.2M and 7.0M, respectively). In addition, CA-MHFA shows notable performance improvements in ER tasks. In the deepfake detection, compared to LLGF, CA-MHFA demonstrates Considerable improvements in most cases. Our results demonstrate consistent improvements across different model scales, highlighting its robustness and versatility.

\section{Conclusion}
In this paper, we propose a novel lightweight back-end module named CA-MHFA for pre-trained SSL-based speaker verification, which can be easily extended to broader speech classification tasks, including emotion recognition and deepfake detection. CA-MHFA enhances contextual modeling by considering frames surrounding the current frame when computing attention weights, leading to more stable and discriminative utterance-level representations. We conducted comprehensive experiments on the VoxCeleb, IEMOCAP, and ASVspoof2019 datasets. The results demonstrate that CA-MHFA outperforms other downstream back-end modules while maintaining lower parameter counts.

\bibliographystyle{IEEEtran}
\bibliography{mybib}
\end{document}